\documentstyle[12pt,aaspp4]{article}

\begin{document}

\title{Interferometric Observations of the T Tauri Stars in the MBM 12 Cloud}
\author{Yoichi Itoh\altaffilmark{1},
Koji Sugitani\altaffilmark{2},
Naoya Fukuda\altaffilmark{3,4},
Kouichiro Nakanishi\altaffilmark{5},
Katsuo Ogura\altaffilmark{6},
Motohide Tamura\altaffilmark{7},
Kazuko Marui\altaffilmark{1}, 
Kenta Fujita\altaffilmark{1},
Yumiko Oasa\altaffilmark{8},
and 
Misato Fukagawa\altaffilmark{9}}

\altaffiltext{1}{Graduate School of Science and Technology, Kobe University,
1-1 Rokkodai, Nada, Kobe, Hyogo 657-8501, Japan, yitoh@kobe-u.ac.jp, fujita, marui@harbor.scitec.kobe-u.ac.jp}
\authoremail{yitoh@kobe-u.ac.jp}
\altaffiltext{2}{Institute of Natural Sciences, Nagoya City University, Mizuho-ku,
Nagoya 467-8501, Japan, sugitani@nsc.nagoya-cu.ac.jp}
\altaffiltext{3}{Japan Science and Technology Corporation, fukudany@sirius.phys.s.chiba-u.ac.jp}
\altaffiltext{4}{Department of Physics, Faculty of Science, Chiba University,
Chikusa-ku, Chiba, 263-8522, Japan}
\altaffiltext{5}{Nobeyama Radio Observatory, Nobeyama, Minamimaki, Minamisaku,
Nagano 384-1305, Japan, nakanisi@nro.nao.ac.jp}
\altaffiltext{6}{Kokugakuin University, 4-10-28 Higashi, Shibuya-ku, 
Tokyo 150-8440, Japan, ogura@kokugakuin.ac.jp}
\altaffiltext{7}{National Astronomical Observatory of Japan, 1-21-1, Osawa, Mitaka, Tokyo 181-8588, Japan, hide@optik.mtk.nao.ac.jp}
\altaffiltext{8}{Earth Observation Research Center, National Space Development Agency of Japan, 1-8-10, Harumi, Chuo-ku, Tokyo, 104-6023, Japan, oasa.yumiko@nasda.go.jp}
\altaffiltext{9}{Department of Astronomy, The University of Tokyo, 7-3-1, Hongo, Bunkyo-ku, Tokyo, 113-0033, Japan, misato@optik.mtk.nao.ac.jp}

\begin{abstract}
We have carried out a millimeter interferometric continuum survey toward 7
YSOs in the MBM 12 cloud.
Thermal emissions associated with 2 YSOs were detected above
the 3-$\sigma$ level at 2.1 mm, and one also showed a 1.3 mm thermal emission.
Another object was marginally detected at 2.1 mm. 
Spectral energy distributions of the YSOs are well fitted by
a simple power-law disk model.
Masses of the circumstellar disks are estimated to be an order of
$0.05 M_{\sun}$.
The circumstellar disks in the MBM 12 cloud have properties in common with
the disks in nearby star-forming regions, in terms of disk
parameters such as a disk mass, as well as an infrared excess.
\end{abstract}


\section{Introduction}

Physical and chemical characteristics as well as evolution process of
a circumstellar disk are tightly connected to formation of
stars and planets.
Millimeter observations, both of continuum emission and of
molecular gas emission, reveal that young stellar objects (YSOs)
commonly have a circumstellar disk (Beckwith et al. 1990).\markcite{Beckwith} 
\markcite{Duvert}Duvert et al. (2000)
 carried out millimeter continuum and line survey for YSOs
in the Taurus molecular cloud. They
detected thermal emission from all classical T Tauri stars (CTTSs) 
in the sample, while they did not detect emission
from any weak-line T Tauri stars (WTTSs)
with a disk mass upper limit of $\sim2\times10^{-4}M_{\sun}$.
They attribute this difference to the evolution
of the circumstellar disks.
Emission from a circumstellar disk, such as
dust continuum emission, molecular
gas emission, and an infrared excess, decrease as the disk evolves. 

The MBM 12 (L 1457) cloud is a high-latitude cloud with a signature of 
star formation. Three LkH$\alpha$ stars and several emission-line
stars have been identified by optical surveys
(e.g. Magnani, Caillault, \& Armus 1990;\markcite{Magnani}
Ogura et al. in preparation)\markcite{Ogura} , an X-ray survey 
(Hearty et al. 2000a),\markcite{HeartyA}
 and  a near-infrared study (Luhman 2001).\markcite{Luhman}
Some sources exhibit mid-infrared excesses, implying circumstellar
disks (Jayawardhana et al. 2001).\markcite{Jaya} 
However, no radio continuum emission nor molecular gas emission
have been detected so far with an upper limit of the
disk mass of $0.09 M_{\sun}$ (Pound 1996;\markcite{Pound}
Hogerheijde et al. 2002).\markcite{Hoge}
\markcite{Hoge}Hogerheijde et al. (2002) claim a deficiency of massive
circumstellar disks in the cloud. If all YSOs in the cloud
do not have even a small disk, the formation of circumstellar disks should
have been prevented, or the disks should have already dissipated.

Until recently the MBM 12 cloud was thought to be one of the nearest
star forming regions with a distance of $\sim$ 65 pc 
(Hobbs, Blitz, \& Magnani 1986;\markcite{Hobbs}
Hearty et al. 2000b).  \markcite{HeartyB}
However, \markcite{Luhman}Luhman (2001) suggests its distance of $\sim$ 275 pc.
\markcite{Andersson}Anderson et al. (2002) claim that there are two components
toward the MBM 12 cloud, with a dense part at $\sim$ 360 pc, and a thin layer
at $\sim$ 80 pc.
If the YSOs are associated with the dense cloud, the YSOs are
low-mass stars with a young age ($\sim$2 Myr), the same generations of or
successors to the YSOs in the nearby star forming regions such as the 
Taurus molecular cloud.
The evolution process of the circumstellar
disks can be discussed by comparing the YSOs in the MBM 12 cloud
with those in the nearby star forming regions.

We present here the results of
a millimeter interferometric continuum survey toward the
YSOs in the MBM 12 cloud. Thanks to high-sensitivity of the Nobeyama
Millimeter Array, we have first detected millimeter continuum emissions from
the YSOs in the MBM 12 cloud.

\section{Observations and Data Reduction}

Radio interferometric observations were carried out in 16 days
in 2001 December, 2002 January, and 2002 December with the Nobeyama 
Millimeter Array. 
The targets are bright YSOs in the MBM 12 cloud identified by the optical
and X-ray wavelengths.
The array consists of six 10 m antennas operating
at rest frequencies of 103.8 GHz ($\lambda = 2.9 $ mm),
141.0 GHz ($\lambda = 2.1 $ mm), and 224.5 GHz ($\lambda = 1.3 $ mm).
The signals from the antennas were
sent to the Ultra Wide Band Correlator (UWBC; Okumura et al. 2000),
which covers 1024 MHz with
a spectral resolution of 8.0 MHz. Total system
temperatures ranged from 200 K to 800 K.
The spatial resolution was about 4$\arcsec$ with the D antenna configuration.
Due to poor weather conditions, the integration time
for 230 GHz observations was strictly limited.
QSO B0234+285 (4C 28.07) was used as the phase calibrator.
The observations were carried out on a cycle of 15 minutes on
the source and 5 minutes on the calibrator.
Total on-source integration times were between 100 minutes and 300 minutes.
The flux of the quasar was calibrated by the observations of Uranus
about every 5 days.
The derived flux have a maximum uncertainty of 20\%.

The data were calibrated and reduced using the UVPROC-II package developed
at the Nobeyama Radio Observatory, and the AIPS package developed at NRAO.
After determining the phase and amplitude gain curve, the uv-visibilities
were gridded and Fourier-transformed using natural weighting to produce
dirty maps with $0\farcs5$ per pixel.
Each dirty map was CLEANed until 1 $\sigma$ noise level.

\section{Results}

\begin{figure}
\plotone{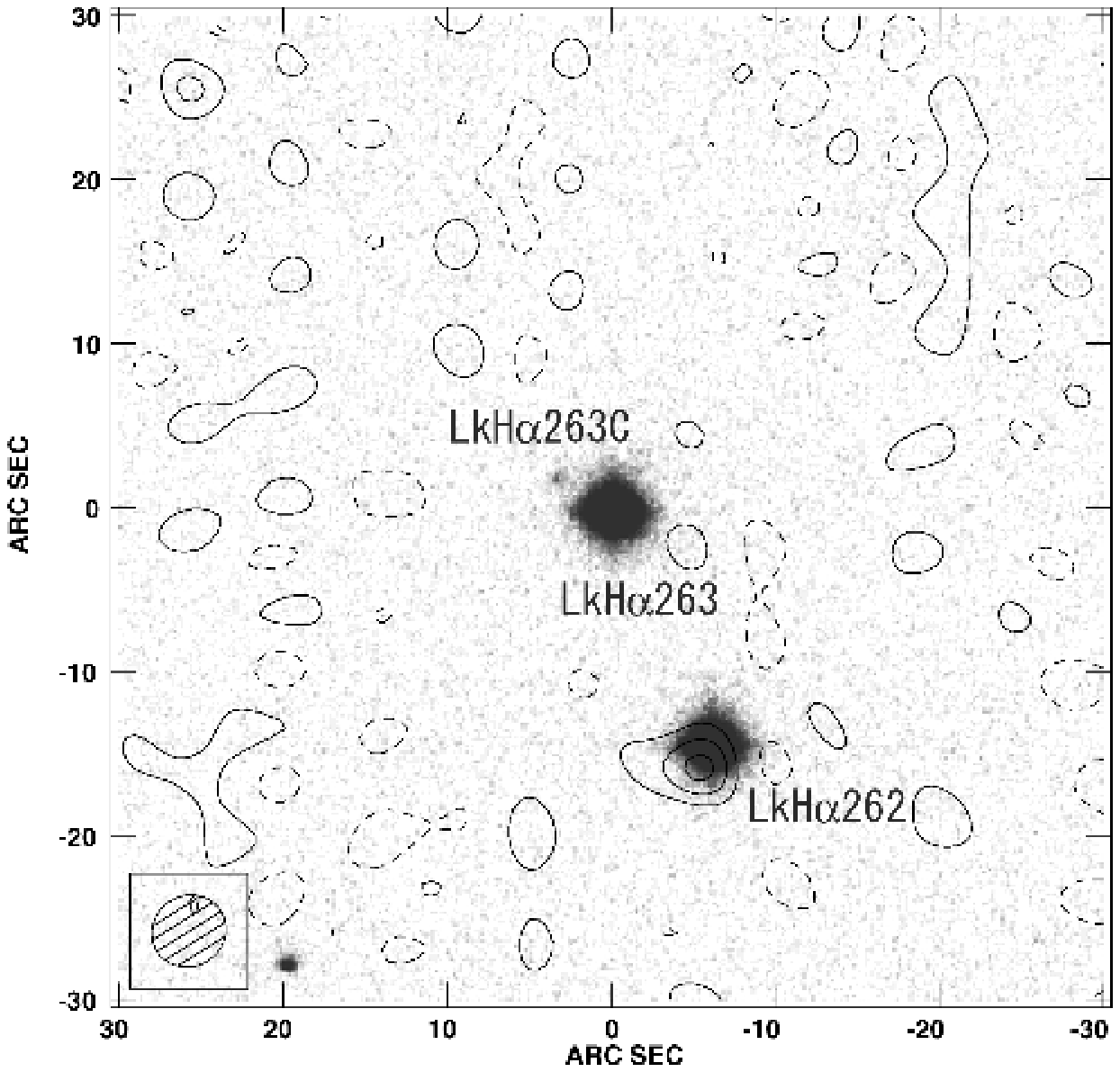}
\caption{Contour of 2.1 mm radio continuum emission toward the LkH$\alpha$ 263
region, overlaid a near-infrared $Kp$-band image taken by the UH 2.2 m 
telescope. 
Contour levels are ($-1.5$, +1.5, +3.0, +4.5) times the noise level (1.75 mJy).
The synthesized beam is shown in the lower left corner of the map.
Primary beam correction is not performed to this image.
Radio continuum emission toward LkH$\alpha$ 262 is clearly detected.
Coordinates are offset to the position of LkH$\alpha$ 263.
\label{Lk262}}
\end{figure}

The survey results are listed in Table \ref{flux-tbl}.
Continuum emissions at 2.1 mm were detected toward LkH$\alpha$ 262 and
LkH$\alpha$ 264 above the 3 $\sigma$ level, and marginally toward
S18. We also detected 1.3 mm continuum emission toward LkH$\alpha$ 264.
At the position of the other objects, no sources were detected above
the rms noise level.
No objects are spatially resolved.
The measured fluxes and the 3 $\sigma$ upper limits of the objects are
summarized in Table \ref{flux-tbl}.
The flux uncertainties for the detected objects include rms of the
sky in the map and a possible 20 \% error in absolute flux
calibrations.
Figure \ref{Lk262} shows a 2.1 mm continuum contour map of the LkH$\alpha$
262/263 region.

\clearpage

\begin{deluxetable}{lccc}
\tablecaption{Observed Flux Densities. \label{flux-tbl}}
\tablehead{
\colhead{Source} 
& \colhead{$F_{2.9 mm}$(mJy)}
& \colhead{$F_{2.1 mm}$(mJy)} 
& \colhead{$F_{1.3 mm}$(mJy)}}
\startdata
LkH$\alpha$ 262  & $<$5.54	& $10.4\pm2.7$ & $<81$ \nl
LkH$\alpha$ 263  & $<$4.82	& $<5$         & $<70$ \nl
LkH$\alpha$ 263C & $<$4.82	& $<5$         & $<70$ \nl
LkH$\alpha$ 264  & $<$10.07	& $15.5\pm3.8$ & $56.2\pm14.5$ \nl
RX J0255.4+2005  & \nodata	& $<12$        & \nodata \nl
E 02553+2018     & \nodata	& $<8$         & \nodata \nl
S18             & \nodata	& $8.0\pm4.6$  & $<$33
\enddata
\end{deluxetable}
\clearpage

\begin{figure}
\plotone{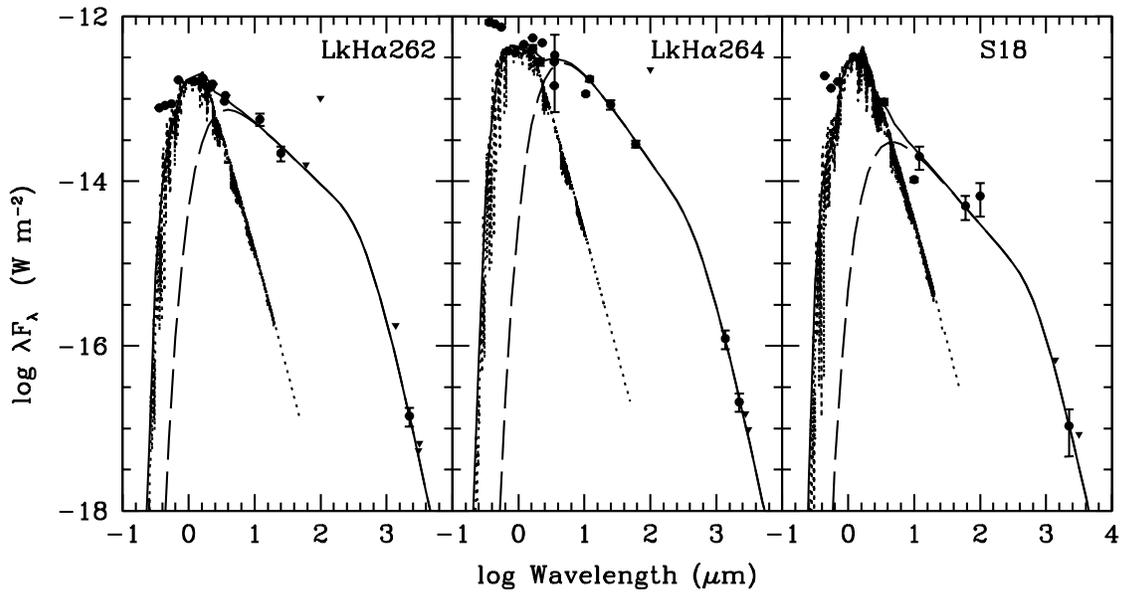}
\caption{Spectral energy distributions of LkH$\alpha$ 262,
LkH$\alpha$ 264, and S18.
Filled circles represent the measured flux values, and filled triangles
represent the upper limits of the measurements.
The dotted line shows the spectral energy distribution of the central star,
and the dashed line shows that of
the circumstellar disk.
The solid line shows the spectral energy distribution of the total flux.
\label{sed}}
\end{figure}

The spectral energy distributions of LkH$\alpha$ 262, LkH$\alpha$ 264, and
S18 are presented
in Figure \ref{sed}.
The data were taken from the 2MASS survey,
the IRAS survey, and previous optical and near-infrared photometric
studies, and mid-infrared photometry (Hearty et al. 2000a;\markcite{HeartyA}
Luhman 2001;\markcite{Luhman} 
Jayawardhana et al. 2001).\markcite{Jaya}
Observed fluxes were dereddened using $A_{V}$ (Luhman 2001)\markcite{Luhman}
with the interstellar extinction law (Rieke \& Lebofsky 1985).\markcite{Rieke}
In these figures, the upper limits are also shown. The spectral energy
distributions of the objects are well fitted
by a combination of a stellar photosphere model 
(Allard, Hauschildt, \& Schweitzer 2000)\markcite{Allard} and 
a power-law disk model (Beckwith et al. 1990).\markcite{Beckwith} 
We adopt a power law form as the disk temperature
$T=T_{1}(r/1 {\rm AU})^{-q}$,
surface density $\Sigma = \Sigma_{1}(r/1 {\rm AU})^{-p}$, and mass opacity
$\kappa_{\nu} = \kappa_{0}(\nu/\nu_{0})^{\beta}$.
At frequencies at which emission from a circumstellar disk are optically
thin, $F_{\nu}\propto\nu^{2+\beta}$. We find $\beta=1.0$ for LkH$\alpha$ 264,
though uncertainties in the millimeter fluxes are very large.
We assume $p=1.0, \kappa_{0}=0.1$ at $\nu_{0} = 10^{12}$ Hz, 
a disk outer radius $R_{D}$ = 100 AU, $\beta = 1.0$, 
and an inclination of $\theta=45^{\circ}$ for all objects.
Note that because none of the objects were spatially resolved, the maximum
outer radius is 550 AU for $d=275$ pc.
The spectral types of the stars are taken from 
\markcite{Luhman}Luhman (2001).
The fitted disk parameters are presented in Table \ref{disk-tbl}. 
Note that the derived disk masses in Table \ref{disk-tbl} can be easily
changed by disk parameters, such as $p$, $\beta$, and $R_{D}$.
The spectral energy distributions of the fitted power-law disk models are
shown in Figure \ref{sed}
by the dashed lines, those
of the stars by the dotted lines, and the composite spectra by the solid lines. 
As seen in the figures, the upper limits of the 3 mm continuum 
observations (Pound et al. 1996; Hogerheijde et al. 2002) are
consistent with the spectral energy distributions derived 
from the models.
Chauvin et al. (2002)\markcite{Chauvin} have
recently discovered an edge-on disk
near LkH$\alpha$ 263 (LkH$\alpha$ 263 C), whose
dust mass is estimated to be  $2\times10^{-6} M_{\sun}$.
This implies that the gas mass of the disk is below
the detection limit of our survey, though this estimated mass is highly 
sensitive to the disk parameters.

\section{Discussion}

The disk masses of the objects are around $0.05
M_{\sun}$, comparable to the intermediate mass disks
in the mearby molecular clouds (e.g. Beckwith et al. 1990;
Andr\'e \& Montlerle 1994).\markcite{Beckwith} 

\clearpage

\begin{deluxetable}{lcccc}
\tablecaption{Disk Parameters. \label{disk-tbl}}
\tablehead{
\colhead{Source} 
& \colhead{$q$} & \colhead{$T_{1}$} & \colhead{$M_{d}$} & \colhead{$r_{0}$} \\
\colhead{} & 
\colhead{} & \colhead{(K)} & \colhead{($M_{\sun}$)} & \colhead{(AU)}}

\startdata
LkH$\alpha$ 262 & 
0.62 & 152 & 0.048 & 0.023 \nl
LkH$\alpha$ 264 & 
0.70 & 220 & 0.085 & 0.080 \nl
S18             & 
0.65 & 112 & 0.071 & 0.024  \nl
\enddata
\end{deluxetable}

\clearpage

Hogerheijde et al. (2002) \markcite{Hoge} observed 7 YSOs in MBM 12 by the
$^{13}$CO (2--1) line. No objects show the molecular line.
Assuming a standard CO abundance,
they estimated upper limits of the disk mass to be 
$(5\sim10)\times10^{-4} M_{\sun}$, which are far smaller than our results
($\sim 0.1 M_{\sun}$).
Therefore, depletion of $^{13}$CO occurs in the disk with 
two orders of magnitude.
Such a depletion is predicted by chemical models (Aikawa et al. 1996)
and is indeed observed commonly in T Tauri stars (e.g. Dutrey, Guilloteau \&
Simon 1994).

\begin{figure}
\plotone{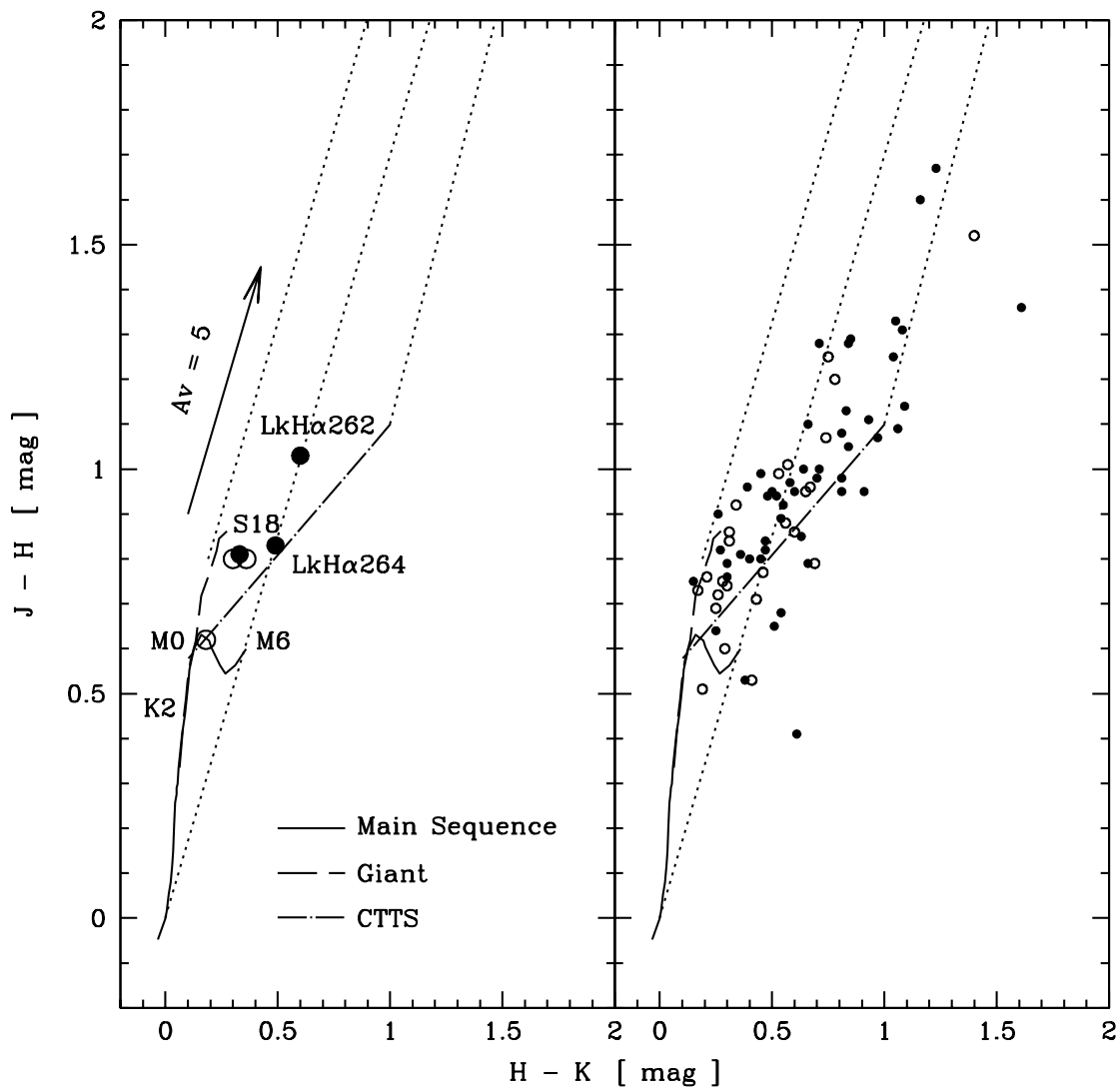}
\caption{ Near-infrared color-color diagrams of the YSOs in the MBM 12 cloud
(left), and in the Taurus molecular cloud listed in Beckwith et al. (1990)
(right).
The objects with radio continuum emission are denoted by the filled circles, 
and the objects without emission by the open circles.
The intrinsic colors of
the main-sequence, giants (Bessell \& Brett 1988), 
and classical T Tauri stars (Meyer, Calvet, \& Hillenbrand 1997) 
are indicated. The reddening vector
follows Koornneef (1983). 
All colors are transformed to the CIT system.
\label{cc}}
\end{figure}

Meyer et al. (1997)\markcite{Meyer} show that a YSO with an accretion
disk has a near-infrared excess.
A near-infrared color-color diagram of the sample is presented in Figure
\ref{cc}. Near-infrared magnitudes are taken from Luhman (2001)
\markcite{Luhman} or measured by us using the UH 2.2 m telescope
(Ogura et al. in preparation).\markcite{Ogura}
As spectral types of LkH$\alpha$ 262 and LkH$\alpha$ 264 are M0 and M3
respectively, they each has a near-infrared excess. Because S18 is of
spectral type of M3, it seems to suffer interstellar reddening without
an intrinsic near-infrared excess. Therefore two of the YSOs with radio
continuum emission
have near-infrared excesses. On the other hand, the YSOs without radio
continuum
emission have near-infrared color consistent with no intrinsic near-infrared
excess. This general trend is also seen in the sample of the
YSOs in Taurus (Figure \ref{cc}).
This can be interpreted as follows: at least two of 
the YSOs with radio continuum emission have an outer portion
of the disk, as well as an inner portion of the disk which generates the
near-infrared excess.

Jayawardhana et al. (2001) \markcite{Jaya} detected near- and mid-infrared
excesses from the YSOs in the MBM 12 cloud. 
All YSOs with radio continuum emission
have the mid-infrared excesses. However it is not the case vice versa.
For example, E 2553+2018 which has the largest $K-L$ excess
and third largest $K-N$ excess in the sample does not have radio 
continuum emission.
N\"urnberger, Chhini, \& Zinnecker (1997) \markcite{Nurnberger} conducted a radio continuum 
survey of T Tauri stars in the Lupus associations.
They find no correlation between the infrared indices (2.2 $\micron$ --
12 $\micron$) and the disk masses.
Also found are no correlations for the YSOs in the Taurus molecular cloud,
the $\rho$ Oph cloud (Andr\'e \& Montmerle 1994) \markcite{Andre}, 
and the Chamaeleon cloud (Henning et al. 1993).\markcite{Henning}
However, there is a trend for the MBM 12 and the Chamaeleon clouds
in which the objects with large radio continuum
emission have large mid-infrared excesses
and the objects without large mid-infrared excesses 
do not have large radio continuum emission.

The YSOs in the MBM 12 cloud are known to have a high binary frequency
(Chauvin et al. 2002). \markcite{Chauvin}
In the sample, LkH$\alpha$ 263, E02553+2018, and S18 are
binaries with small separations, whereas the other objects are single stars or
a wide binary. However the small number of the sample prevents us to 
investigate the relationship between a binary and a circumstellar disk.

The circumstellar disks in the MBM 12 cloud detected by this survey have
properties in common with the disks in nearby star-forming regions, 
in terms of disk parameters, such as a disk mass, as well as an infrared excess.

\acknowledgments

We owe great thanks to the staff members of the Nobeyama Radio Observatory for
their help during the observations and data reduction.
Y.I. is supported by the Sumitomo foundation.
K. N. is supported financially by the JSPS fellowship.

\clearpage

%

\end{document}